\documentclass[a4paper,10pt,eqno]{article}
\usepackage{theorem}
\usepackage{latexsym,amssymb,amsfonts,amsmath}
\usepackage{cite}
\newtheorem{thm}{Theorem}[section]

\newtheorem{cor}[thm]{Corollary}

\theoremheaderfont{\scshape}
\newcommand{\RM}{\mathbb{R}}
\newcommand{\ZM}{\mathbb{Z}}

\newcommand{\CM}{\mathbb{C}}

\newcommand{\ket}[1]{|#1\rangle}

\title{{\Large {\bf Entanglement for discrete-time quantum walks \\
on the line}}}
\author{
{\small Yusuke Ide\footnote{To whom correspondence should be addressed. E-mail: ide@kanagawa-u.ac.jp}}\\
{\scriptsize Department of Information Systems Creation, 
Faculty of Engineering, 
Kanagawa University}\\
{\scriptsize Kanagawa, Yokohama 221-8686, Japan}\\
{\scriptsize e-mail: ide@kanagawa-u.ac.jp}\\
{\small Norio Konno}\\
{\scriptsize Department of Applied Mathematics, 
Faculty of Engineering, 
Yokohama National University}\\
{\scriptsize Hodogaya, Yokohama 240-8501, Japan}\\
{\scriptsize e-mail: konno@ynu.ac.jp}\\
{\small Takuya Machida}\\
{\scriptsize Meiji Institute for Advanced Study of Mathematical Sciences,  
Meiji University}\\
{\scriptsize Tama, Kawasaki 214-8571, Japan}\\
{\scriptsize e-mail: bunchin@meiji.ac.jp}\\
}
\date{\empty }
\begin{document}
\maketitle

\par\noindent
\begin{small}
\par\noindent
{\bf Abstract}. The discrete-time quantum walk is a quantum counterpart of the random walk. It is expected that the model plays important roles in the quantum field. In the quantum information theory, entanglement is a key resource. We use the von Neumann entropy to measure the entanglement between the coin and the particle's position of the quantum walks. Also we deal with the Shannon entropy which is an important quantity in the information theory. In this paper, we show limits of the von Neumann entropy and the Shannon entropy of the quantum walks on the one dimensional lattice starting from the origin defined by arbitrary coin and initial state. In order to derive these limits, we use the path counting method which is a combinatorial method for computing probability amplitude. 
\footnote[0]{
{\it Abbr. title:} Entanglement for quantum walks
}
\end{small}

\setcounter{equation}{0}

\section{Introduction}
The discrete-time (or coined) quantum walk (QW) has been extensively studied by many authors as a quantum analogue of the random walk \cite{Kempe2003,Kendon2007,VAndraca2008,Konno2008b,AharonovEtAl2001,AmbainisEtAl2001}. As the random walk plays important roles in various fields, it is expected that the QW also plays such roles in the quantum field. In fact, benefits of using QWs have been shown in various applications, for example, quantum speed-up algorithm \cite{Ambainis2003,ChildsEtAl2003,Ambainis2004,ShenviEtAl2003} and universal quantum computation \cite{Childs2009,LovettEtAl2010}. 

In this paper, we give asymptotic behaviors of the von Neumann entropy and the Shannon entropy of discrete-time QWs on $\ZM$, where $\ZM$ is the set of integers. Because entanglement does not appear in classical systems, it is an important concept for quantum information processing (see Nielsen and Chuang \cite{NielsenChuang2000}). We use the von Neumann entropy of the reduced density matrix of the coin to quantify the entanglement between the coin and the particle's position. The Shannon entropy is one of a basic quantity in the information theory and it clarifies information included in the system. In the present paper, we calculate the entropies of the QWs starting from the origin with arbitrary coin and initial state from a path counting approach which is a combinatorial method for computing probability amplitude. 
By numerical simulations, Carneiro et al. \cite{CarneiroEtAl2006} studied the long-time asymptotic coin-position entanglement (aCPE) of QWs on various graphs, for examples, $\ZM, \ZM^2,$ triangular lattices, cycles. Venegas-Andraca and Bose \cite{VenegasBose2009} also investigated the von Neumann entropy on $\ZM$ numerically. Using Fourier analysis techniques, aCPE of the Hadamard walk on $\ZM$ for both localized (i.e., our setting) and non-localized initial conditions was analytically computed by Abal et al. \cite{AbalEtAl2006}. In the similar technique, Annabestani et al. \cite{AnnabestaniEtAl2009} gave an exact characterization of the aCPE of QWs on $\ZM^2$. The evolution is determined by the tensor product of two one-qubit Hadamard operations. Liu and Petulante \cite{LiuPetulante2010} presented limit theorems for the von Neumann entropy of QWs on the $N$-cycle. 
Bracken et al. \cite{BrackenEtAl2004} numerically computed the Shannon entropy of the QW on $\ZM$ defined by a coin which is a generalization of the Hadamard coin. The numerical result of Chandrashekar et al. \cite{ChandrashekarEtAl2008} suggests that several properties of the Shannon entropy the QW on $\ZM$ given by another generalization of the Hadamard coin. 

The rest of this paper is organized as follows. In Sect. 2, we give the definition of the QW. Results on the von Neumann entropy are presented in Sect. 3. Section 4 is devoted to the proof of our main result (Theorem {\rmfamily \ref{momijiman}}). In the final section, a long-time asymptotic for the Shannon entropy of the QW is shown.

%彼等の 2 次元の解析はかなりハードに行われているように見えるが，
%もっと一般のモデルでの解析は今後のテーマとなり得るかもしれない．
%グローヴァーや離散フーリエ変換など
%Venegas-Andraca and Bose \cite{VenegasBose2009} の扱いをどうするか．

\section{Definition}
The discrete-time QW is a quantum counterpart of the classical random walk with additional degree of freedom called chirality. The chirality takes values left and right, and it means the direction of the motion of the walker. At each time step, if the walker has the left chirality, it moves one step to the left, and if it has the right chirality, it moves one step to the right. Let define
\begin{eqnarray*}
\ket{L} = 
\left[
\begin{array}{cc}
1 \\
0  
\end{array}
\right],
\qquad
\ket{R} = 
\left[
\begin{array}{cc}
0 \\
1  
\end{array}
\right],
\end{eqnarray*}
where $L$ and $R$ refer to the left and right chirality state, respectively.  

Let $\hbox{U}(2)$ denote the set of $2 \times 2$ unitary matrices. The time evolution of the QW on $\ZM$ is determined by 
\begin{eqnarray*}
U =
\left[
\begin{array}{cc}
a & b \\
c & d
\end{array}
\right] \in \hbox{U}(2),
\end{eqnarray*}
with $a, b, c, d \in \CM$ and $\CM$ is the set of complex numbers. The unitarity of $U$ gives 
\begin{eqnarray*}
|a|^2 + |b|^2 =|c|^2 + |d|^2 =1, \> a \overline{c} + b \overline{d}=0, \> c= - \triangle \overline{b}, \> d= \triangle \overline{a},
\label{konno-eqn:seisitu}
\end{eqnarray*}
where $\overline{z}$ is the complex conjugate of $z \in \CM$ and $\triangle = \det U = a d - b c$ with $|\triangle|=1.$ In particular, we write 
\begin{eqnarray*}
U (\theta) =
\left[
\begin{array}{cc}
\cos \theta & \sin \theta \\
\sin \theta & - \cos \theta
\end{array}
\right],
\end{eqnarray*}
where $0 < \theta < \pi/2$. When $\theta = \pi/4$, the QW is called the Hadamard walk.

In order to define the dynamics of the model, we divide $U$ into two matrices:
\begin{eqnarray*}
P =
\left[
\begin{array}{cc}
a & b \\
0 & 0 
\end{array}
\right], 
\quad
Q =
\left[
\begin{array}{cc}
0 & 0 \\
c & d 
\end{array}
\right],
\end{eqnarray*}
with $U=P+Q$. The matrix $P$ (resp. $Q$) represents that the walker moves to the left (resp. right) at each time step. 
Let $\Xi_{n}(l, m)$ denote the sum of all paths starting from the origin in the trajectory consisting of $l$ steps left and $m$ steps right. In fact, for time $n = l+m$ and position $x=-l + m$, we have 
\begin{align*}
\Xi_n (l,m) = \sum_{l_j, m_j} P^{l_{1}} Q^{m_{1}} P^{l_{2}} Q^{m_{2}} \cdots P^{l_{n-1}} Q^{m_{n-1}} P^{l_{n}} Q^{m_{n}},
\end{align*}
where the summation is taken over all integers $l_j, m_j \ge 0$ satisfying $l_1+ \cdots +l_n=l, \> m_1+ \cdots + m_n = m, \> l_j+ m_j=1$. We should note that the definition gives 
\begin{align*}
\Xi_{n+1}(l, m) = P \> \Xi_{n}(l-1, m) + Q \> \Xi_{n}(l, m-1).
\end{align*}
For example, in the case of $l=3, \> m=1$, we have 
\begin{align*}
\Xi_4 (3,1) &= QP^3 + PQP^2 + P^2QP + P^3 Q. 
%\label{kaede}
\end{align*}

The set of initial qubit states at the origin for the QW is given by 
\begin{eqnarray*}
\Phi = \left\{ \varphi =
\alpha \ket{L}+\beta \ket{R} \in \mathbb C^2 :
|\alpha|^2 + |\beta|^2 =1
\right\},
\end{eqnarray*}
where $T$ is the transposed operator. The probability that a quantum walker is in position $x$ at time $n$ starting from the origin with $\varphi \in \Phi$ is defined by 
\begin{align*}
P (X_{n} =x) = || \Xi_{n}(l, m) \varphi ||^2,
\end{align*}
where $n=l+m$ and $x=-l+m$. The probability amplitude $\Psi_n (x)$ at position $x$ and time $n$ is given by
\begin{align}\label{eqpsin}
\Psi_n (x) =  \Psi_n^L (x)\ket{L}+\Psi_n^R (x)\ket{R} = \Xi_{n}(l, m) \varphi.
\end{align}
So $P (X_{n} =x) = |\Psi_n^L (x)|^2 + |\Psi_n^R (x)|^2.$

Let $|\Psi_n \rangle = {}^T [ \ldots, \Psi_n (-1), \Psi_n (0), \Psi_n (1), \ldots ].$ The density operator at time $n$ is given by $\rho_n = | \Psi_n \rangle \langle \Psi_n |$. Entanglement for pure states can be quantified by the von Neumann entropy of the reduced density operator $\rho_n ^c = \hbox{Tr}_p \> (\rho_n)$, where the partial trace is taken over position. The associated von Neumann entropy at time $n$ is 
\begin{align*}
S_n ^c = - \hbox{Tr} \> \left\{ \rho_n ^c \> \log_2 (\rho_n ^c) \right\}.
\end{align*}
This is known as entropy of entanglement, which quantifies the quantum correlations present in the pure state. Let 
\begin{align*}
|| \Psi_n^{L} ||^2 &= \sum_{x=-n}^{n} \left| \Psi_n^{L} (x) \right|^2, \quad 
|| \Psi_n^{R} ||^2 = \sum_{x=-n}^{n} \left| \Psi_n^{R} (x) \right|^2, \quad \\
\langle \Psi_n^{R}, \Psi_n^{L} \rangle &= \sum_{x=-n}^{n} \overline{\Psi_n^{R} (x)} \Psi_n^{L} (x). 
\end{align*}
Note that $|| \Psi_n^{L} ||^2 + || \Psi_n^{R} ||^2=1$ for any time $n$. 
The entropy of entanglement can be obtained after diagonalization of $\rho_n ^c$. This is represented by the following $2 \times 2$ Hermitian matrix 
\begin{align*}
\rho_n ^c 
=
%\left[
%\begin{array}{cc}
%\rho_n ^{L} & \rho_n ^{RL} \\
%\overline{\rho_n ^{RL}} & \rho_n ^{R} 
%\end{array}
%\right]
%=
\left[
\begin{array}{cc}
|| \Psi_n^{L} ||^2 & \langle \Psi_n^{R}, \Psi_n^{L} \rangle \\
\langle \Psi_n^{L}, \Psi_n^{R} \rangle &  || \Psi_n^{R} ||^2
\end{array}
\right].
\end{align*}
Let 
\begin{align*}
\triangle_n (\rho) = \det (\rho_n ^c ) = || \Psi_n^{L} ||^2 || \Psi_n^{R} ||^2 - |\langle \Psi_n^{L}, \Psi_n^{R} \rangle|^2.
\end{align*}
Then the reduced entropy is expressed as 
\begin{align*}
S_n ^c = - \left( r_{n,+} \log_2 (r_{n,+}) + r_{n,-} \log_2 (r_{n,-}) \right),
\end{align*}
where $r_{n, \pm}$ are eigenvalues of $\rho_n ^c$ and given by 
\begin{align*}
r_{n,\pm} = \frac{1 \pm \sqrt{1 - 4 \triangle_n (\rho)}}{2}.
\end{align*}
Here $0 \log_2 0$ is interpreted as $0$. The determinant $\triangle_n (\rho) \in [0, 1/4]$ quantifies the coin-position entanglement in the QW. The greater the value of $\triangle_n (\rho)$, the greater the entanglement. If $\triangle_n (\rho) = 0$, i.e., a product state, then $S_n ^c =0$. If $\triangle_n (\rho) = 1/4$, i.e., a maximally coin state, then $S_n ^c =1$. We will compute the asymptotic value of $S_n ^c$ as time $n \to \infty$ for the QW determined by $U \in \hbox{U}(2)$ with initial qubit state $\varphi \in \Phi$.

On the other hand, the entropy of the reduced density matrix of the position at time $n$, $\rho_n ^p  = \hbox{Tr}_c \> (\rho_n)$, also quantifies the entanglement between the coin and the walker's position. This is represented by $\infty \times \infty$ Hermitian matrix whose $(x,y)$ element, $\rho_n ^p (x,y),$ is  
\begin{align*}
\rho_n ^p (x,y) = \Psi_n^{L} (x) \overline{\Psi_n^{L}(y)} + \Psi_n^{R} (x) \overline{\Psi_n^{R}(y)},
\end{align*}
where $x, \> y \in \ZM.$ The diagonal element becomes 
\begin{align*}
\rho_n ^p (x,x) = \left| \Psi_n^{L} (x) \right|^2 + \left| \Psi_n^{R} (x) \right|^2 = P (X_n =x).
\end{align*}
Note that if $|x| > n$ or $|y| > n$, then $\rho_n ^p (x,y) = 0$. Therefore the reduced entropy, $S_n ^p$, can be expressed as 
\begin{align*}
S_n ^p = - \sum_{x=-n} ^{n} r_{n} (x) \log_2 \left( r_{n} (x) \right),
\end{align*}
where $r_{n} (x)$ are eigenvalues of $(2n+1) \times (2n+1)$ Hermitian matrix with the element $\rho_n ^p (x,y)$ for $x, \> y \in \{-n, -n+1, \ldots , n \}$. We should remark that $S_n ^c = S_n ^p$ for any $n \ge 0$, so we focus on $S_n ^c$.

\section{Results on entropy of entanglement}
In this section we present the following main result of aCPE for the QW on $\ZM$. Put 
\begin{align*}
|| \Psi_{\infty}^{L} || = \lim_{n \to \infty} || \Psi_n^{L} ||, \quad || \Psi_{\infty}^{R} || = \lim_{n \to \infty} || \Psi_n^{R} ||, \quad \langle \Psi_{\infty}^{R}, \Psi_{\infty}^{L} \rangle = \lim_{n \to \infty} \langle \Psi_n^{R}, \Psi_n^{L} \rangle.
\end{align*}
\begin{thm}
\label{momijiman}
When the QW is determined by $U$ with $abcd \not= 0$, we have
\begin{align*}
|| \Psi_{\infty}^{L} ||^2
&=
\left( 1 - \frac{|b|}{2} \right) \> | \alpha |^2 + \frac{|b|}{2} \> | \beta |^2 + \frac{\Gamma}{2(1+|b|)}, 
%\label{man1}
\\ 
|| \Psi_{\infty}^{R} ||^2 
&=
\frac{|b|}{2} \> | \alpha |^2 + \left( 1 - \frac{|b|}{2} \right) \> | \beta |^2 - \frac{\Gamma}{2(1+|b|)}, 
%\label{man2}
\\
\langle \Psi_{\infty}^{R}, \Psi_{\infty}^{L} \rangle
&=
\frac{|b|}{2a \overline{b}} \> \left\{ |b|(1- |b|) \left(| \alpha |^2 - | \beta |^2 \right) + \frac{|b| \> \Gamma + a \alpha \overline{b \beta} - \overline{a \alpha} b \beta}{1+|b|} \right\},
%\label{man3}
%\\
%\lim_{n \to \infty} \sum_{x=-n}^{n} \Psi_n^{R} (x) \overline{\Psi_n^{L} (x)} 
%&=
%\frac{|b|}{2 \overline{a} b} \> \left[ |b|(1- |b|) 
%\left(| \alpha |^2 - | \beta |^2 \right) + \frac{|b| \> \Gamma 
%+ \overline{a \alpha} b \beta - a \alpha \overline{b \beta}}{1+|b|} \right].
\end{align*}
where $\Gamma = a \alpha \overline{b \beta} + \overline{a \alpha} b \beta.$ Then 
\begin{align*}
\lim_{n \to \infty} S_n ^c = - \left\{ r_{\infty,+} \log_2 (r_{\infty,+}) + r_{\infty,-} \log_2 (r_{\infty,-}) \right\},
\end{align*}
where
\begin{align*}
r_{\infty,\pm} = \frac{1 \pm \sqrt{1 - 4 \triangle_{\infty} (\rho)}}{2}.
\end{align*}
and
\begin{align*}
\triangle_{\infty} (\rho) = || \Psi_{\infty}^{L} ||^2 || \Psi_{\infty}^{R} ||^2 - |\langle \Psi_{\infty}^{L}, \Psi_{\infty}^{R} \rangle|^2.
\end{align*}
\end{thm}
In particular, if $U=U(\theta)$, then the following result can be obtained. 
\begin{cor}
\label{momijicor}
When the QW is determined by $U(\theta)$ with $0 < \theta < \pi/2$, we have 
\begin{align}
|| \Psi_{\infty}^{L} ||^2
&=
\left( 1 - \frac{\sin \theta}{2} \right) \> | \alpha |^2 + \frac{\sin \theta}{2} \> | \beta |^2 + \frac{\sin \theta (1 - \sin \theta)}{2 \cos \theta} \> \left( \alpha \overline{\beta} + \overline{\alpha} \beta \right), 
\label{mancor1}
\\ 
|| \Psi_{\infty}^{R} ||^2 
&=
\frac{\sin \theta}{2} \> | \alpha |^2 + \left( 1 - \frac{\sin \theta}{2} \right) \> | \beta |^2 - \frac{\sin \theta (1 - \sin \theta)}{2 \cos \theta} \> \left( \alpha \overline{\beta} + \overline{\alpha} \beta \right), 
\label{mancor2}
\\
\langle \Psi_{\infty}^{R}, \Psi_{\infty}^{L} \rangle 
&=
\frac{\sin \theta (1 - \sin \theta )}{2 \cos \theta} \> \left\{ | \alpha |^2 - | \beta |^2  + \frac{\sin \theta \> \left( \alpha \overline{\beta} + \overline{\alpha} \beta \right) + \alpha \overline{\beta} - \overline{\alpha} \beta}{\cos \theta} \right\}. 
\nonumber
%\label{mancor3}
\end{align}
\end{cor}
Reversible cellular automata were considered in \cite{KonnoEtAl2004} by the Fourier analysis. As a special case, the automaton includes the QW defined by $U(\theta)$. Eqs. (\ref{mancor1}) and (\ref{mancor2}) are equivalent to equations given in pp.416--417 of \cite{KonnoEtAl2004}. The correspondence is $|| \Psi_n ^L || \leftrightarrow || \Psi^L (n)||, \> || \Psi_n ^R || \leftrightarrow || \Psi^R (n)||, \> \alpha \leftrightarrow \alpha_l, \> \beta \leftrightarrow \alpha_r$. Here we use a different approach based on a path counting. Abal et al. \cite{AbalEtAl2006} investigated the case of $U(\pi/4)$ (i.e., Hadamard walk) for the localized initial condition (i.e., our setting) and the non-localized initial condition in the position space spanned by $|\pm 1 \rangle$. Noting that their notation $r_{1,2}$ corresponds to our notation $r_{\infty ,\pm}$, their result for the localized case of $\varphi = {}^T [\cos \theta_1, e^{i \theta_2} \sin \theta_1] \in \Phi$ (stated at Appendix in \cite{AbalEtAl2006}) is consistent with Corollary {\rmfamily \ref{momijicor}}. As for the QW on the line starting at the origin, Carneiro et al. \cite{CarneiroEtAl2006} studied the QW defined by $U$ with $a=-d=\sqrt{\rho}, \> b=c=\sqrt{1-\rho}$ ($\rho=0.5$ case is the Hadamard walk) and the initial condition $\varphi = {}^T [\cos \theta_1, e^{i \theta_2} \sin \theta_1] \in \Phi$. They numerically found that the entanglement oscillates around an asymptotic value and the rate of convergence depends on the symmetry in the distribution of the QW. More symmetric distributions about the origin converge faster than asymmetric ones. Their limiting value for the Hadamard case is consistent with Corollary {\rmfamily \ref{momijicor}}.  
Recently, All\'es et al. \cite{AllesEtAl2010} studied an entanglement among two walkers on the line. We note that the model can be exactly mapped to our model by the following correspondence in their notations:
\begin{align*}
P_{\text{(left move)}} 
&\Rightarrow 
\begin{bmatrix}
0 & 0\\
-\beta ^{*}\sqrt{\rho }-\alpha ^{*}\sqrt{1-\rho }e^{-i(\theta +\eta )} & -\beta ^{*}\sqrt{1-\rho }e^{i(\theta -\eta )}+\alpha ^{*}\sqrt{\rho }e^{-2i\eta }
\end{bmatrix},\\
Q_{\text{(right move)}} 
&\Rightarrow 
\begin{bmatrix}
\alpha \sqrt{\rho }-\beta\sqrt{1-\rho }e^{-i(\theta +\eta )} & \alpha \sqrt{1-\rho }e^{i(\theta -\eta )}+\beta \sqrt{\rho }e^{-2i\eta }\\
0 & 0
\end{bmatrix}.
\end{align*}

%(これが厳密に示せるであろうか？)

\section{Proof of Theorem {\rmfamily \ref{momijiman}}}
In this section we assume $abcd \not=0$. We consider the following four matrices: 
\begin{eqnarray*}
P =
\left[
\begin{array}{cc}
a & b \\
0 & 0 
\end{array}
\right], 
\quad
Q =
\left[
\begin{array}{cc}
0 & 0 \\
c & d 
\end{array}
\right],
\quad 
R =
\left[
\begin{array}{cc}
c & d \\
0 & 0 
\end{array}
\right], 
\quad
S =
\left[
\begin{array}{cc}
0 & 0 \\
a & b 
\end{array}
\right]. 
\end{eqnarray*}
Put $x \wedge y = \min \{x,y \}$. For $l \wedge m \ge 1$, we have 
\begin{align*}
\Xi_n (l,m) 
= a^l \overline{a}^m \triangle^m \sum_{\gamma =1}^{l \wedge m} \left(- \frac{|b|^2}{|a|^2} \right)^{\gamma} {l-1 \choose \gamma-1} {m-1 \choose \gamma -1} \left( \frac{l - \gamma}{a \gamma} P + \frac{m - \gamma}{\triangle \overline{a} \gamma} Q - \frac{1}{\triangle \overline{b}} R + \frac{1}{b} S\right), 
\end{align*}
by the path counting method \cite{Konno2002, Konno2005}. Therefore, by noting Eq.\ (\ref{eqpsin}), we have
\begin{align*}
\Psi_n^{L} (x) 
&= a^l \overline{a}^m \triangle^m \sum_{\gamma =1}^{l \wedge m} \left(- \frac{|b|^2}{|a|^2} \right)^{\gamma} {l-1 \choose \gamma-1} {m-1 \choose \gamma -1} \left( \frac{l - \gamma}{a \gamma} A - \frac{1}{\triangle \overline{b}} B \right), 
\\
\Psi_n^{R} (x) 
&= a^l \overline{a}^m \triangle^m \sum_{\gamma =1}^{l \wedge m} \left(- \frac{|b|^2}{|a|^2} \right)^{\gamma} {l-1 \choose \gamma-1} {m-1 \choose \gamma -1} \left( \frac{m - \gamma}{\triangle \overline{a} \gamma} B + \frac{1}{b} A \right), 
\end{align*}
where $A = a \alpha + b \beta, \> B = c \alpha + d \beta.$ Let $[x]$ denote the integer part of $x \in \RM$. We first consider $||\Psi_{\infty}^{L}||$ case. For $1 \le l \le [n/2]$, 
\begin{align*}
|\Psi_n^{L} (x)|^2 
&= |a|^{2n} \sum_{\gamma =1}^{l} \sum_{\delta =1}^{l} \left(- \frac{|b|^2}{|a|^2} \right)^{\gamma + \delta} {l-1 \choose \gamma-1} {l-1 \choose \delta -1} {n-l-1 \choose \gamma-1} {n-l-1 \choose \delta -1} \\
\times & \left[ \frac{l^2 |A|^2}{\gamma \delta |a|^2} - \frac{l}{\gamma} \left\{ \frac{|A|^2}{|a|^2} + \Theta \right\} - \frac{l}{\delta} \left\{ \frac{|A|^2}{|a|^2} + \overline{\Theta} \right\}  + \left\{ \frac{|A|^2}{|a|^2} + \frac{|B|^2}{|b|^2} + 2 \Re (\Theta) \right\} \right],
\end{align*}
where $\Theta = (A \overline{B})/(\overline{\triangle}ab).$ Furthermore we will rewrite $|\Psi_n^{L} (x)|^2$ by using Jacobi polynomials. Let $P^{\nu, \mu} _n (x)$ denote the Jacobi polynomial which is orthogonal on $[-1,1]$ with respect to $(1-x)^{\nu}(1+x)^{\mu}$ with $\nu, \mu > -1$. Then the following relation holds:
\begin{align*}
P^{\nu, \mu} _n (x) = \frac{\Gamma (n + \nu + 1)}{\Gamma (n+1) \Gamma (\nu +1)} \> {}_2F_1(- n, n + \nu + \mu +1; \nu +1 ;(1-x)/2),
%\label{konno-eqn:ken}
\end{align*}
where ${}_2F_1(a, b; c ;z)$ is the hypergeometric series and $\Gamma (z)$ is the gamma function. In general, as for orthogonal polynomials, see \cite{Andrews1999}. Then we have
\begin{align}
\sum_{\gamma =1} ^{l}
\left(- \frac{|b|^2}{|a|^2} \right)^{\gamma -1}
{1 \over \gamma} 
{l-1 \choose \gamma- 1}  
{n-l-1 \choose \gamma- 1} 
&= 
\frac{|a|^{-2(l-1)}}{l} P^{1,n-2l} _{l-1}(2|a|^2-1), 
\label{yukari1}
\\
\sum_{\gamma =1} ^{l}
\left(- \frac{|b|^2}{|a|^2} \right)^{\gamma -1}
{l-1 \choose \gamma- 1}  
{n-l-1 \choose \gamma- 1} 
&= |a|^{-2(l-1)} P^{0,n-2l} _{l-1}(2|a|^2-1).
\label{yukari2}
\end{align}
By using Eqs. (\ref{yukari1}) and (\ref{yukari2}), we see that for $1 \le l \le [n/2]$, 
%\begin{lem}
%\label{sirowine}
\begin{align*}
|\Psi_n^{L} (x)|^2 
&= |a|^{2n-4l} |b|^4 \left[ \frac{|A|^2}{|a|^2} (P^1)^2  - 2 \left\{ \frac{|A|^2}{|a|^2} + \Re (\Theta) \right\} (P^1 P^0) \right.
\\
& \qquad \qquad \qquad \qquad \qquad \qquad \qquad \left. + \left\{ \frac{|A|^2}{|a|^2} + \frac{|B|^2}{|b|^2} + 2 \Re (\Theta) \right\} (P^0)^2\right], 
\end{align*}
where $P^{i} = P^{i,n-2l} _{l-1}(2|a|^2-1) \> (i=0,1)$. 
%\end{lem}
By a similar argument in \cite{Konno2002, Konno2005}, we obtain
\begin{align}
\lim_{n \to \infty} ||\Psi_n^{L}||^2 =  \lim_{n \to \infty} \sum_{x=-n}^{n} \left| \Psi_n^{L} (x) \right|^2 = \frac{|b|^3}{2\pi} \int_{-|a|}^{|a|} \frac{h^L(x) }{\sqrt{|a|^2-x^2}} \> dx, 
\label{akiba}
\end{align}
where
\begin{align*}
h^L(x) = \frac{|A|^2}{|a|^2|b|^2} \> \frac{1-x}{1+x} -2 \left\{ \frac{|A|^2}{|a|^2} + \Re (\Theta) \right\} \> \frac{1}{1+x} + \left\{ \frac{|A|^2}{|a|^2} + \frac{|B|^2}{|b|^2} + 2 \Re (\Theta) \right\},
\end{align*}
and $\Re (z)$ is the real part of $z \in \CM$. 
Indeed, if $n\to \infty $ with $k/n\in (-(1-|a|)/2, (1+|a|)/2)$, then 
\begin{align*}
P^{0}&\sim \frac{2|a|^{2k-n}}{\sqrt{\pi n\sqrt{-\Lambda }}}\cos(An+B), \\
P^{1}&\sim \frac{2|a|^{2k-n}}{\sqrt{\pi n\sqrt{-\Lambda }}}\sqrt{\frac{x}{(1-x)(1-|a|^2)}}\cos(An+B+\theta ),
\end{align*}
where $\Lambda =(1-|a|^2)\{(2x-1)^2-|a|^2\}$, $A$ and $B$ are some constants which are independent of $n$, and $\theta \in [0,\pi/2]$ is determined by $\cos \theta =\sqrt{(1-|a|^2)/4x(1-x)}$. By these asymptotics and the Riemann-Lebesgue lemma, we have Eq.\ (\ref{akiba}). 

Noting that
\begin{align*}
&\int_{-|a|}^{|a|} \frac{dx}{\sqrt{|a|^2-x^2}} 
= \pi, \qquad  
\int_{-|a|}^{|a|} \frac{1}{1+x} \> \frac{dx}{\sqrt{|a|^2-x^2}} = \frac{\pi}{\sqrt{1-|a|^2}} = \frac{\pi}{|b|}, \quad 
\\
&\int_{-|a|}^{|a|} \frac{1-x}{1+x} \> \frac{dx}{\sqrt{|a|^2-x^2}} 
= \frac{(2-\sqrt{1-|a|^2})\pi}{\sqrt{1-|a|^2}} = \frac{(2 - |b|) \pi}{|b|},
\end{align*}
we have
\begin{align*}
\lim_{n \to \infty} ||\Psi_n^{L}||^2 = \frac{(2-|b|)|A|^2}{2} + \frac{|b||B|^2}{2} + |b|^2(|b|-1) \Re (\Theta).
\end{align*}
Moreover, the following relations hold: 
\begin{align*}
|A|^2 
&= |a|^2 |\alpha|^2 + |b|^2 |\beta|^2 + \Gamma, \quad |B|^2 = |b|^2 |\alpha|^2 + |a|^2 |\beta|^2 - \Gamma, 
\\
\Re (\Theta) 
&= |\beta|^2 - |\alpha|^2 + \frac{(|a|^2 - |b|^2) \Gamma}{2 |a|^2 |b|^2}, 
\end{align*}
where $\Gamma = a \alpha \overline{b \beta} + \overline{a \alpha} b \beta.$ Therefore we obtain the desired conclusion. Concerning $||\Psi_{\infty}^{R}||$, we get the following result by a similar fashion:
\begin{align*}
\lim_{n \to \infty} ||\Psi_n^{R}||^2 = \frac{|b|^3}{2\pi} \int_{-|a|}^{|a|} \frac{h^R(x) }{\sqrt{|a|^2-x^2}} \> dx, 
\end{align*}
where
\begin{align*}
h^R(x) = \frac{|B|^2}{|a|^2|b|^2} \> \frac{1+x}{1-x} -2 \left\{ \frac{|B|^2}{|a|^2} - \Re (\Theta) \right\} \> \frac{1}{1-x} + \left\{ \frac{|B|^2}{|a|^2} + \frac{|A|^2}{|b|^2} - 2 \Re (\Theta) \right\}.
\end{align*}
So the desired conclusion is obtained. Finally we consider $\langle \Psi_{\infty}^{R}, \Psi_{\infty}^{L} \rangle$ case. For $1 \le l \le [n/2]$, 
\begin{align*}
&\overline{\Psi_n^{R} (x)} \Psi_n^{L} (x) \\
&= |a|^{2n} \sum_{\gamma =1}^{l} \sum_{\delta =1}^{l} \left(- \frac{|b|^2}{|a|^2} \right)^{\gamma + \delta} {l-1 \choose \gamma-1} {l-1 \choose \delta -1} {n-l-1 \choose \gamma-1} {n-l-1 \choose \delta -1} \\
&\times \left[ \frac{l(n-l) A \overline{B}}{\gamma \delta \overline{\triangle} a^2} - \frac{l}{\gamma} \left( \frac{A \overline{B}}{\overline{\triangle} a^2} - \frac{|A|^2}{a \overline{b}} \right) - \frac{n-l}{\delta} \left( \frac{A \overline{B}}{\overline{\triangle} a^2} + \frac{|B|^2}{a \overline{b}} \right) + \frac{A \overline{B}}{\overline{\triangle} a^2} - \frac{\overline{A}B}{\triangle \overline{b}^2} +\frac{|B|^2 - |A|^2}{a \overline{b}} \right].
\end{align*}
Hence we have
\begin{align*}
\lim_{n \to \infty}  \langle \Psi_n^{R}, \Psi_n^{L} \rangle = \frac{|b|^3}{2\pi} \int_{-|a|}^{|a|} \frac{h^{RL} (x) }{\sqrt{|a|^2-x^2}} \> dx, 
\end{align*}
where
\begin{align*}
h^{RL} (x) 
&= \frac{A \overline{B}}{\overline{\triangle} a^2 |b|^2} - \left( \frac{A \overline{B}}{\overline{\triangle} a^2} - \frac{|A|^2}{a \overline{b}} \right) \> \frac{1}{1+x} 
\\
&\qquad \qquad \qquad 
- \left( \frac{A \overline{B}}{\overline{\triangle} a^2} + \frac{|B|^2}{a \overline{b}} \right) \> \frac{1}{1-x} + \frac{A \overline{B}}{\overline{\triangle} a^2} - \frac{\overline{A}B}{\triangle \overline{b}^2} +\frac{|B|^2 - |A|^2}{a \overline{b}}.
\end{align*}
This proves the case of $\langle \Psi_n^{R}, \Psi_n^{L} \rangle$. So the proof of Theorem {\rmfamily \ref{momijiman}} is complete.

\section{Results on the Shannon entropy}
In this section, we compute the Shannon entropy of the QW. Let $P_n (x) = P (X_{n} =x)$. Then we define the Shannon entropy of the QW by
\begin{align*}
S_n =  - \sum_{x=-n}^n \> P_n (x) \log_2  P_n (x).
\end{align*}
%When $P_n (x)$ is a uniform distribution on $W_n = \{-n, -(n-2), \ldots, n\}$, i.e., $P_n (x) = 1/(n+1)$ for any $x \in W_n$, then $S_n$ takes the maximum value 
%\begin{align*}
%S_n ^{\max} = \log_2 (n+1).
%\end{align*}
%Furthermore, for 
Furthermore, we let 
\begin{align*}
S_n ^{D} = - \sum_{x=-n}^{n} \frac{|\Psi_n^{D}(x)|^2}{|| \Psi_n^{D} ||^2} \> \log_2 \left( \frac{|\Psi_n^{D}(x)|^2}{|| \Psi_n^{D} ||^2} \right),
\end{align*}
for $D \in \{L,R\}$. 
In this section, we present limit theorems of the Shannon entropy for $S_n ^L, S_n ^R,$ and $S_n$. Define
\begin{align}\label{eqdens}
\rho^D = \lim_{n \to \infty} || \Psi_n^{D} ||^2, 
\end{align}
and
\begin{align*}
f^{D}(x) = \frac{|b|^3}{2 \pi} \> \frac{h^{D}(x)}{\sqrt{|a|^2 - x^2}},
\end{align*}
with $D \in \{L, R\}$. Moreover
\begin{align*}
f(x) = f^{L}(x) + f^{R}(x) = \frac{|b|}{\pi (1-x^2) \sqrt{|a|^2 - x^2}} \left\{   1 - \left( |\alpha|^2 - |\beta|^2 + \frac{\Gamma}{|a|^2} \right) x \right\}.
\end{align*}
The following long-time behavior of the Shannon entropy is shown. 
\begin{thm}
\label{momijimanj}
If the QW is determined by $U$ with $abcd \not= 0$, then we have
\begin{align} 
\lim_{n \to \infty} \frac{S_n^D}{\log_2 (n/2)} = \lim_{n \to \infty} \frac{S_n}{\log_2 (n/2)} = 1, 
\label{manjj}
\end{align}
where $D \in \{L, R\}$. Furthermore, 
\begin{align} 
\lim_{n \to \infty} \log_2 (n/2) \left( \frac{S_n^D}{\log_2 (n/2)} -1 \right) 
&= -\int_{-|a|}^{|a|} \frac{ f^{D}(x) }{ \rho^{D} } \log_2 \left( \frac{ f^{D}(x) }{ \rho^{D} } \right) dx,
\label{manj1}
\\
\lim_{n \to \infty} \log_2 (n/2) \left( \frac{S_n}{\log_2 (n/2)} -1 \right) 
&= -\int_{-|a|}^{|a|} f (x) \log_2 \left( f(x) \right) dx,
\label{manj2}
\end{align}
where $D \in \{L, R\}$. 
\end{thm}
Bracken et al. \cite{BrackenEtAl2004} numerically computed the Shannon entropy of the QW defined by $U$ with $a=-d=\sqrt{p}, \> b=c=\sqrt{1-p}$ ($p=0.5$ case is the Hadamard walk). The numerical result of Chandrashekar et al. \cite{ChandrashekarEtAl2008} suggests that the Shannon entropy $S_n$ of the QW given by $U (\theta) \> (0 \le \theta \le \pi/2)$ is maximum for $\theta = \pi/4$ (the Hadamard walk), and $S_n (\theta_1) \le S_n (\theta_2)$ if $0 \le \theta_1 \le \theta_2 \le \pi/4$ and $S_n (\theta_1) \ge S_n (\theta_2)$ if $\pi/4 \le \theta_1 \le \theta_2 \le \pi/2$ for any $n$. 
Our result is consistent with the numerical results. Moreover, we give the limiting values of the Shannon entropies explicitly which are new findings of this article. 

When $P_n (x)$ is a uniform distribution on $W_n = \{-n, -(n-2), \ldots, n\}$, i.e., $P_n (x) = 1/(n+1)$ for any $x \in W_n$, then $S_n$ takes the maximum value 
\begin{align*}
S_n ^{\max} = \log_2 (n+1).
\end{align*}
In Eq. (\ref{manjj}) we can replace $\log_2 (n/2)$ by $S_n ^{\max}$, since $S_n ^{\max} = \log_2 (n+1).$ 

For the symmetric random walk whose walker moves one unit to the right with probability $1/2$ and to the left with probability $1/2$, the following result of the Shannon entropy, $S_n^{RW}$, is known, (see \cite{Barron1986, Takano1987} for more detailed information): 
\begin{align*} 
\lim_{n \to \infty} \frac{S_n^{RW}}{\log_2 (\sqrt{n})} = 1.
%\label{manjjj}
\end{align*}
Furthermore, 
\begin{align*} 
\lim_{n \to \infty} \log_2 (\sqrt{n}) \left( \frac{S_n^{RW}}{\log_2 (\sqrt{n})} -1 \right) = \frac{1}{2} \> \log_2 \left( 2 \pi e \right).
%\label{manjj1}
\end{align*}
This shows that the rate of convergence of $S_{n}^{RW}$ is related to that of 
walker's position in the central limit theorem (CLT). 
Eqs. (\ref{manjj}), (\ref{manj1}) and (\ref{manj2}) shows that in the QW case, 
the rate of convergence of the Shannon entropy is also related to that of the
CLT. But the relativity is slightly different from the random walk case. 
\\

%Venegas-Andraca and Bose \cite{VenegasBose2009} では，
%基本的に $P_n (x)$ の代わりに $\left| \Psi_n^{L} (x) \right|^2$ 
%と $\left| \Psi_n^{R} (x) \right|^2$ を計算している． 

\par\noindent
{\it Proof.} We here consider only $S_n^L$ case. Other cases, i.e., $S_n^R$ and $S_n$, can also be shown in a similar way. $S_n ^{L}$ is rewritten as  
\begin{align*}
S_n ^{L} = \log_2 \left( || \Psi_n^{L} ||^2 \right) - \frac{1}{|| \Psi_n^{L} ||^2} \sum_{x=-n}^{n} |\Psi_n^{L}(x)|^2 \> \log_2 \left( |\Psi_n^{L}(x)|^2 \right).
\end{align*}
By Eq. (\ref{eqdens}), we have
\begin{align*}
\lim_{n \to \infty} S_n ^{L} = \log_2 \left( \rho^L \right) - \frac{1}{\rho^L} \lim_{n \to \infty} \sum_{x=-n}^{n} |\Psi_n^{L}(x)|^2 \> \log_2 \left( |\Psi_n^{L}(x)|^2 \right).
\end{align*}
As in a similar proof of Theorem {\rmfamily \ref{momijiman}}, we obtain 
\begin{align*}
&\sum_{x=-n}^{n} |\Psi_n^{L}(x)|^2 \> \log_2 \left( |\Psi_n^{L}(x)|^2 \right) 
\\
&\sim \int_{-|a|}^{|a|} f^{L}(x) \log_2 \left( 2 f^{L}(x) /n \right) dx 
\\
&= -\int_{-|a|}^{|a|} f^{L}(x) dx \times \log_2 (n/2)  + \int_{-|a|}^{|a|} f^{L}(x) \log_2 \left( f^{L}(x) \right) dx
\\
&= -\rho^L \times \log_2 (n/2)  + \int_{-|a|}^{|a|} f^{L}(x) \log_2 \left( f^{L}(x) \right) dx,
\end{align*}
where $a(n) \sim b(n)$ means $a(n)/b(n) \to 1$ as $n \to \infty$. The third equality comes from Eq. (\ref{akiba}). Thus
\begin{align*}
S_n ^{L} \sim \log_2 (n/2) - \int_{-|a|}^{|a|} \frac{ f^{L}(x) }{ \rho^{L} } \log_2 \left( \frac{ f^{L}(x) }{ \rho^{L} } \right) dx,  
\end{align*}
and this completes the proof.

\section{Summary}
In this paper, we show  limit theorems for the von Neumann entropy and the Shannon entropy of the QWs on $\ZM$ starting from the origin with arbitrary coin and initial state. In order to compute the entropies, we use a path counting method which is a powerful tool for analysis of QWs on $\ZM$. By using this method, we can derive limiting values of the entropies explicitly. But in general, it is hard to use the path counting method for higher dimensional cases. For the RW case, the Shannon entropy plays an important role in the theory of large deviation principle. It is a natural question what is a role of the entropy in the large deviation principle for the QW. 

\par
\
\par\noindent
{\bf Acknowledgments.} 
%The author thanks T. Namiki, K. Nakamura, T. Machida for valuable comments. 
This work was partially supported by the Grant-in-Aid for Scientific Research (C) of Japan Society for the Promotion of Science (Grant No. 21540118).

%\par
%\
%\par

\begin{small}

\end{small}

\end{document}